\documentclass[12]{iopjournal}

\usepackage{graphicx}
\usepackage{dcolumn}
\usepackage{bm}
\usepackage{amsmath}
\usepackage{amsthm}
\usepackage{amssymb}
\usepackage{bbm}
\usepackage{braket}
\usepackage{nicefrac}
\usepackage{mathrsfs}
\usepackage{color}
\usepackage{physics}
\usepackage{ulem}
\usepackage{enumerate}
\usepackage{verbatim}
\usepackage{hyperref}

\hypersetup{
	colorlinks = true,
	urlcolor   = blue,
	linkcolor  = blue,
	citecolor  = cyan
}

\newcommand{\ketup}{\ket{\uparrow}}
\newcommand{\ketdown}{\ket{\downarrow}}
\newcommand{\coinoperator}{\hat{\mathcal{C}}}
\newcommand{\shiftoperator}{\hat{\mathcal{S}}}
\newcommand{\stepoperator}{\hat{\mathcal{U}}}

\newcommand{\hilbertspace}{\mathscr{H}}

\newcommand{\coinspace}{\hilbertspace_c}
\newcommand{\positionspace}{\hilbertspace_p}

\newcommand{\probdistr}{\mathbb{P}}
\newcommand{\shannonentropy}{H}

\newcommand{\mustar}[2]{\mu^{\star #1}(#2)}

\newtheorem{theorem}{Theorem}[section]
\newtheorem{corollary}{Corollary}[theorem]

\begin{document}
\articletype{Article type}

\title{Quantum-driven  sampling of the almost-uniform distribution via  quantum walks  }

\author{Marco~Radaelli$^{1,2,3}$,Claudia~Benedetti$^{1,4,*}
$, Stefano~Olivares$^{1,4,*}$}

\affil{$^1$Dipartimento di Fisica ``Aldo Pontremoli'', Universit\`a degli Studi di Milano, via~Celoria~16, I-20133 Milan, Italy}
\affil{$^2$School of Physics, Trinity College Dublin, Dublin 2, Ireland}\\
\affil{$^3$Trinity Quantum Alliance, Unit 16, Trinity Technology and Enterprise Centre, Pearse Street, Dublin 2, D02 YN67, Ireland}\\
\affil{$^4$Istituto Nazionale di Fisica Nucleare, Sezione di Milano, I-20133 Milan, Italy}

\affil{$^*$Author to whom any correspondence should be addressed.}
\\
\email{claudia.benedetti@unimi.it - stefano.olivares@fisica.unimi.it}

\begin{abstract}
We investigate the use of discrete-time quantum walks   to sample from an almost-uniform distribution, in the absence of any external source of randomness. 
Integers are encoded on the vertices of a cycle graph, and a quantum walker  evolves for a fixed number of steps before its position is measured and recorded. 
The walker is then reset to the measured site, and the procedure is iterated to generate the sequence of outcomes.
We show that {when the number of nodes of the graph is odd,} i.e. the condition of the ergodic theorem for {classical} random walks on finite groups are satisfied,  the marginal distributions converges   to the uniform distribution.
Although correlations between successive outcomes are unavoidable, they can be significantly reduced by a suitable choice of the evolution time.
\end{abstract}

\section*{Introduction}

Uniform distributions are central to many applications in science and technology, since  they are the fundamental  model for processes with equally likely outcomes. In random number generation, they are essential for producing unbiased random sequences and they are  the basis to derive more complex distributions. 
Indeed, many technological applications require random numbers, such as  Monte Carlo simulations, cryptography, numerical integration and testing of computer programs \cite{knuth_2020}. 
In general, we can identify two main families of methods to generate random numbers. On the one hand, \textit{pseudo-random} number generators exploit deterministic algorithms, able to provide sequences of values having similar statistical properties to a truly random sequence \cite{frederick20}. 
On the other hand, \textit{true-random} number generators are based on either chaotic classical physical systems \cite{LEcuyer_2017,torregrosa19} or quantum mechanical devices \cite{ma16,Herrero_Collantes_2017,Mannalath22}. 
Among the most used methods to produce a string of random numbers there are  radioactive decay \cite{Manelis1961,Vincent70} and quantum optical systems \cite{Jennewein2000,Furst10,ren11,collins15}.
In most random number generation tasks, it is essential to produce values sampled from a uniform distribution, which can then be used either  directly or as a building block for generating other distributions. 

Recently, discrete-time quantum walks (DTQWs)  have been used to develop random number generator (RNG) protocols  \cite{Sarkar_2019,bae21,bae22}. 
DTQWs  were first introduced as one of the possible ways to naturally generalize classical random walks into a quantum mechanical framework \cite{Aharonov_1993}, together with their continuous-time counterparts \cite{Fahri_1998,frigerio21}. 
Although the relation between the two models has not been completely clarified, it has been subject of intense recent research 
\cite{strauch06,childs10,dalessandro2010,Schmitz16}.
A DTQW describes the coherent motion of a quantum particle that can jump among connected discrete positions in discrete time steps and {is equipped with } an internal degree of freedom, called the quantum coin. \cite{Kempe_2003,Venegas-Andraca_2012}.  

DTQWs find applications in different contexts. They provide a universal model for quantum computation \cite{lovett10}   and they are employed to solve  a variety of problems, such as  quantum spatial search \cite{shenvi2003,Potocek09,lovett19}, quantum teleportation \cite{Wang17,Yamagami:2021}  graph isomorphism \cite{Douglas_2008,liu19} and quantum metrology \cite{annabestani22,cavazzoni24,cavazz24}.   DTQWs are also used to achieve quantum  state transfer \cite{Kurzynski11,zhan14,stefan16}. 
Moreover, they are suitable candidates to build  cryptographic protocols \cite{vlachou15, vlachou18},
 security schemes \cite{rohde12,Chandrashekar2015} and image encryption \cite{yang15}.
{Wavefunction interference in the unitary DTQW dynamics leads to spatial probability distributions that depend sensitively on the initial state, the coin operator, and the evolution time. When measurements are performed, the Born rule maps the quantum state to intrinsically random classical outcomes, which can be exploited for randomness generation } \cite{ abdellatif2019, ellatif2020}. 
\newline
\newline
{Even though techniques for randomness generation are well established, we consider a scenario in which the only available physical resource is a  quantum walk and we investigate whether randomness  can be generated from its dynamics, analyzing the strengths and limitations of the resulting approach.}
{In particular, we analyze whether it is possible to  exploit a DTQW to generate outcomes whose probability distribution is uniform over the sample space. This means that  
each outcome has a uniform marginal distribution. 
Accordingly, in this work we do not address the problem of random number generation, but we  focus specifically on the task of uniform sampling in this sense.}

We focus on DTQW on cycle graphs, which, compared to quantum walks on an infinite line, allow for a compact implementation and, thus, can be embedded in more complex networks. 
Suitable platforms to implement these systems include  optical \cite{schreiber12,Smirne_2020,segawa22},  photonic  \cite{Rohde_2011,qiang24,morandotti25} and   solid-state physics settings \cite{ryan05,Manouchehri_2008,ramasesh17}; moreover, they can also be efficiently simulated on  quantum computers and processors \cite{razzoli24, gong2}.

Here, we first briefly review  previously proposed algorithms \cite{Richter_2007,Sarkar_2019} and then we introduce {a  protocol to uniform sampling}, based on the iterated convolution of the transition probability of a DTQW. 
{Specifically, the resulting dynamics can be described in terms of classical Markov chains driven by transition probabilities derived from the discrete-time quantum walk dynamics.}
We also assess the presence of correlations in the sequence of  drawn numbers which are typical of this kind of protocol, and we provide an operational  strategy to reduce them.

The paper is organized as follows: In Sect.~\ref{sect:Discrete_time_quantum_walks} we introduce the basic tools to describe  discrete-time quantum walks together with the concept of randomness associated with its limiting distribution;
in Sect.~\ref{sect:approaches} we 
briefly review existing  DTQW-based protocols for almost uniform sampling and, then, in Sect.~\ref{sect:sampling} we present a novel scheme for sampling numbers from a uniform distribution on a cycle graph using DTQW, along with a proof of its convergence.
We conclude the paper with some final remarks in Sect.~\ref{sect:Conclusions}.

\section{Discrete-time quantum walks}\label{sect:Discrete_time_quantum_walks}
Discrete-time quantum walks, first introduced 
in Rqf.~\cite{Aharonov_1993}, 
describe the  time-step evolution of a quantum particle with an internal degree of freedom, called quantum coin, on a  set of $N$ discrete positions.
They 
are defined on the composite  Hilbert space $\coinspace\otimes\positionspace$, where $\positionspace$ represents a positional Hilbert space, while $\coinspace$ is the {coin} space. 
The spatial basis vectors are localized states over the $N$ discrete positions  $\{\ket{x}\} \in \positionspace $, with $x=0,\ldots,N-1$. 
The coin is a two-level quantum system  described by the orthonormal basis  $\{\ketup$, $\ketdown\}$  in $\coinspace$.
 The unitary operator $\stepoperator$, that is applied at each time-step,  is composed by two contributions:
a coin-flip operator $\coinoperator$, acting on $\coinspace$,  and a conditional shift operator 
$\shiftoperator$ on the global system. 
The coin operator  $\coinoperator $  {can be written as a general element of $U(2)$: 
\begin{align}
\coinoperator(\theta,\alpha,\beta,\gamma)=
e^{i\gamma}\begin{pmatrix}
\cos\theta\, e^{i\alpha} & \sin\theta\, e^{i\beta} \\
-\sin\theta\, e^{-i\beta} & \cos\theta\, e^{-i\alpha}
\end{pmatrix},
\end{align}
which includes the Hadamard coin
$\mathcal{C}_H=\coinoperator\left(\frac{\pi}{4},\frac{\pi}{2},\frac{\pi}{2},-\frac{\pi}{2}\right)$
and the $x$-axis rotation $\coinoperator_X=\coinoperator\left(\frac{\pi}{4},0,\frac{\pi}{2},0
    \right)$.
}
%

We consider discrete spatial positions arranged as the vertices of a cycle graph, with the boundary condition $\ket{N}\equiv\ket{0}$. 
The conditional shift  $\shiftoperator$ then moves the walker, according to the state of the coin, between adjacent sites:
    \begin{align}
        \shiftoperator =  \ketbra{\uparrow}{\uparrow} &\otimes \sum_{x=0}^{N-1} \ketbra{(x+1)_{N}}{x}+ \ketbra{\downarrow}{\downarrow}\otimes \sum_{x=0}^{N-1} \ketbra{(x-1)_N}{x},
    \end{align}
where the symbol $(x)_N$ stands for $x\!\!\mod  N$.
The single-step operator $\stepoperator$ is therefore defined as:
\begin{equation}
    \label{stepoperator}
    \stepoperator = \shiftoperator \left(\coinoperator\otimes\hat{\mathbb{I}}_p\right),
\end{equation}
where $\hat{\mathbb{I}}_p$ is the identity operator on the positional space.
Given an initial state $\ket{\Psi_0}\in \coinspace\otimes \positionspace$, the final state of the DTQW after $t\in\mathbb{N}$ steps is given by
\begin{equation}
    \ket{\Psi_t} = \stepoperator^t \ket{\Psi_0}.
    \label{evol}
\end{equation}
The probability for the walker to be found on site $\ket{x}$ after $t$ steps is
\begin{equation}
p_x(t)=|\bra{\uparrow}\otimes\braket{x}{\Psi_t}|^2+|\bra{\downarrow}\otimes\braket{ x }{\Psi_t}|^2.
\label{bornrule}
\end{equation}

The outcomes of a  position measurement for  the walker 
are intrinsically random, distributed according to the probability distribution  in Eq.~\eqref{bornrule}. 
This randomness can be exploited to generate strings of  random numbers that can be used also for quantum cryptographyc tasks. Therefore, 
it is important to have a tool to assess  the
unpredictability of the QW measurement outcomes.

Entropy is a reliable way  to quantify the randomness of independent and identically distributed outcomes. Consider a random variable $X$  with possible discrete values $(x_0,x_1,\dots,x_{N-1})$ and  probability distribution $\probdistr(X)=(p_0,p_1,\dots,p_{N-1})$, where $p_k$ is the probability to obtain the outcome $x_k$, $k=0,\ldots\,N-1$.  
The randomness associated with the random variable $X$ can be computed with the Shannon entropy \cite{shannon48}, defined as:
\begin{align}
   & \shannonentropy(X) = -\sum_{x=0}^{N-1} p_x \log_2 p_x.
\end{align}
In this framework, the set $\{x_k\}$ corresponds to the possible QW positions over a cycle graph and the set $\{p_k\}$ to the QW spatial probability distribution  at time $t$, as represented by Eq.~\eqref{bornrule}.

When generating randomness, { the goal is to produce outcomes whose probability distribution has maximal  entropy.} The maximum value of the Shannon entropy is obtained for a variable $X_u$ distributed according to the uniform distribution $\probdistr_u(X_u)=(1/N, 1/N,\dots,1/N)$ and it takes the value $H(X_u)=\log_2 N$. The uniform distribution  thus  serves as the reference  for randomness generation.
{It is therefore necessary to  quantify how close a given probability distribution is to the uniform one. To this purpose, we employ  the Kolmogorov distance, also known as the total variation distance, between two discrete probability distributions $\{p_x\}$ and $\{q_x\}$,  \cite{nielsen_chuang_2010}:
\begin{align}
K\left(\{p_x\},\{q_x\}\right)=\frac12 \sum_{x=0}^{N-1}\,\left|p_x-q_x\right|.
\label{kolmogorov}
\end{align}
The quantity in Eq. \eqref{kolmogorov} vanishes if and only if the two distributions are identical.
}
{Following Refs.~\cite{Richter_2007},  {\it almost-uniform sampling} refers to a procedure that outputs  elements whose distributions are within $\epsilon$ total variation distance of the uniform distribution over a finite set, for an arbitrary accuracy parameter 
$\epsilon>0$.
}

\section{Approaches to  (alomost) uniform randomness via QW }\label{sect:approaches}
\begin{figure}[t]
    \centering   \includegraphics[width=0.5\linewidth]{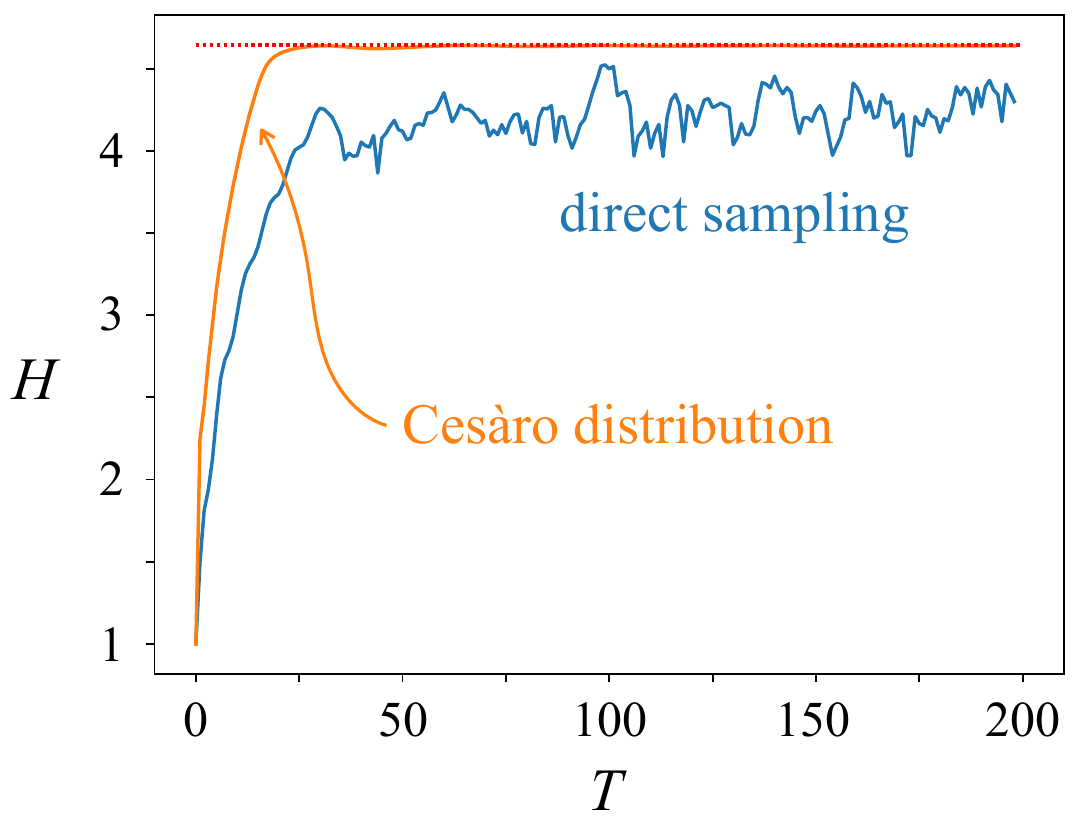}
    \caption{Shannon entropy  of the spatial probability distribution  on an $N=25$ cycle as a function of timesteps $T$, both for the direct sampling and for the Cesàro distribution obtained with Hadamard coin (see the main text for details).}
    \label{fig:Cesàro_distribution}
\end{figure}

In this section, we briefly review two techniques that have been employed to achieve an approximate uniform sampling through quantum walk measurements on a cycle graph.
\newline
The simplest technique consists of sampling directly from the spatial distribution of the quantum walker after a fixed number of timesteps $T$. 
Due to the reversible nature of the quantum evolution, this distribution does not converge to the uniform one, regardless of the number of timesteps considered.
Therefore, an approximation is necessary, and 
$T$ must be chosen to minimize the deviation of the distribution from uniformity, corresponding to the maximization of its Shannon entropy.
In Fig.~\ref{fig:Cesàro_distribution},  the behavior of the Shannon entropy $H$ of the QW-distribution (direct sampling) is shown as a function of the number of timesteps $T$, compared with the uniform one.
To achieve quasi-uniformity 
$T$ must be chosen to correspond to the maximum value of 
$H$ within the attainable timestep interval. 

The random number generation performance of quantum walks on the bi-infinite line by using this direct protocol has been discussed in Ref.~\cite{Sarkar_2019}.
\newline
\newline
A different technique is based on the sampling of the so called Cesàro distribution~\cite{Aharonov_2001}. This protocol introduces an element of irreversibility in the procedure, thus allowing the convergence of the spatial distribution to the almost uniform one.

Given the  probability distribution of the position variable $X$, $\probdistr_t(X)$ at the timestep $t$, the Cesàro distribution is defined as its time-average up to  $T\gg1$ timesteps, namely:
\begin{equation}
    \overline{\probdistr}_{T}(X) = \frac{1}{T}\sum_{t=1}^T \probdistr_t(X).
\end{equation}
It is straightforward to prove that, $\forall T$, $\overline{\probdistr}_{T}(X)$ is a normalized probability distribution if and only if $\probdistr_t(X)$ is. 
If the number of positions on the cycle~$N$ is odd then the time-average distribution, obtained using the Hadamard coin, converges to the uniform one (see Ref.~\cite{Aharonov_2001}), as shown in Fig.~\ref{fig:Cesàro_distribution}. 
However, if the evolution operator \eqref{stepoperator} admits degeneracy in the spectrum, the Cesàro distribution 
does not converge to the uniform one. 
Specifically, for the coin operator  $\coinoperator_X$, the uniformity cannot be attained (see   Appendix~\ref{app:Cesàro_convergence} for details). 

An efficient protocol for sampling from the Cesàro distribution was proposed in Ref.~\cite{Richter_2007} and consists of the following steps: first, a random integer time $t\in [0,T]$ is generated; the quantum walk is then evolved up to time $t$, after which a position measurement is performed; these steps are repeated as many times as needed to obtain  a sample of the desired length.
This protocol allows for the sampling of an  asymptotically uniform distribution.
However, a significant limitation  is that it requires as input a source of uniform randomness, to be used to extract the  times $t$ (for details about this protocol and its performance see Ref.~\cite{Richter_2007}).

\section{ 
Almost-Uniform Sampling via DTQW }
\label{sect:sampling}
In this section, we propose an alternative method for almost-uniform sampling based on the evolution of a discrete-time quantum walk on a cycle graph. Unlike the approach of Ref.~\cite{Richter_2007}, here we assume that we have no access to an external source of randomness. It relies on the spatial measurements of a DTQW  and  the concept of convolution of probability distributions.\\
Due to the translational symmetry of the cycle graph, the protocol is independent of the choice of the initial position $x_0$.
The steps of the protocol are the following:
\begin{enumerate}[I.]
    \item fix an integer number  $m$ and prepare the initial state $\ket{\Psi_0}$ for the DTQW by choosing an arbitrary initial position $x_0$ and setting the coin state {to an arbitrary fixed initial state} $\ket{c_0}$,
    such that $\ket{\Psi_0}=\ket{c_0}\otimes \ket{x_0}$;
    \item evolve the quantum walk for $m$ steps, according to Eq. \eqref{evol}; then perform a position measurement and record the outcome $x$;
    \item reset the coin to its initial state and initialize the walker at the measured position $x$, i.e., $\ket{\Psi_0}=\ket{c_0}\otimes \ket{x}$. 
    \item repeat steps II-III 
    as many times as needed to generate a sequence of integers. 
\end{enumerate}
{The protocol generates a sequence of outcomes whose extracted values are marginally uniformly distributed in the asymptotic regime. This marginal uniformity, however, does not imply that the outcomes are independent and identically distributed (i.i.d.), since the underlying process is Markovian rather than i.i.d. }
Notably, re-initializing the walker at the previously measured position (see the step III) introduces correlations among successive  outputs, as each new initial state depends on the outcome of the previous measurement.
This distinguishes the protocol from the one in Section \ref{sect:approaches}, which sampled i.i.d. random variables. Below, we discuss the implications of these correlations for random number generation. 
Here we just note that, if access to an external source of randomness is allowed, such that it is possible to extract a random number $q$ uniformly in $[0,N-1]$, then one could replace the initial state in step III above with $\ket{c_0}\otimes \ket{q}$.
In this way, one can still get the smoothening effect of the change in the initial position, but without introducing correlations.
However, using an external source of randomness is not always feasible, and it is precisely this scenario that we are addressing in the present work.

With the introduction of the correlations mentioned above, it is very important to clarify the terminology that we are going to employ. In particular, the  protocol  generates a realization of a string of  correlated random variables $X_{1:S} =\{ X_1\ldots X_S$\}, associated with a \textit{joint} probability distribution $\probdistr(X_{1:S})$. 
The marginal probability distribution $\probdistr(X_n)$ refers to the $n$-th  outcome  alone, and it is obtained by marginalising over all the other random variables in the string.

In the present protocol, the probability distribution of the variable $X_n$  only depends on the  previously extracted value $X_{n-1}$,  the dependence being given by the spatial re-initialization of the  walker. 
The joint probability distribution  $\probdistr(X_{1:S})$ then corresponds to  a {classical} Markov chain, where the transition probabilities are generated by the very quantum dynamics of the walk after $m$ steps:
\begin{align}
\label{condprob}
    \probdistr(&X_n=x_n|X_{n-1}=
    x_{n-1})=
    \sum_c|\langle c\otimes x_n|\hat{\mathcal{U}}^m\ket{c_0\otimes x_{n-1}}|^2
\end{align}
in accordance with Eq.~\eqref{bornrule}. 
\newline
In the following, we  prove that the marginal distribution $\probdistr(X_n)$, {namely that of  the $n^{\rm{th}}$  outcome, } converges to the uniform distribution in the asymptotic regime  $n\gg1$. 

\subsection{Convergence of the marginal $\probdistr(X_n)$ to the uniform distribution}\label{ss:convergence}
The discrete rotational symmetry of the cycle implies invariance under translations of the initial node of the walk and therefore   we can choose site $x_0=0$ as the starting point, without loss of generality. 
The transition probability for the first step of the protocol,  which we call $\mu(x_1)$, is the conditional probability of reaching node $x_1$ starting from the initial node $0$:
\begin{align}
 & \mu(x_1)=\probdistr(X_1=x_1|X_0=0).
\end{align}
{$\mu(x)$ is also referred to as the propagator and, for a fixed graph size $N$, it depends on the coin operator, the number of steps $m$ and the  initial state of the system $\ket{\Psi_0}$, in accordance with Eq. \eqref{condprob}.  } 
Hereafter, we use the shorthand notations:
\begin{align}
\probdistr(X_j|X_k)&\equiv \probdistr(X_j=x_j|X_k=x_k)\\
\probdistr(X_j)&\equiv \probdistr(X_j=x_j).
\end{align}
The  rotational symmetry of the cycle graph implies that 
\begin{align}
\probdistr(X_2|X_1)=\probdistr(X_2\ominus X_1|X_0=0)=\mu(x_2\ominus x_1),
\end{align}
where the symbol $\ominus$ represents the difference modulo $N$. In other words, the probability of finding the quantum walker at position $x_2$ starting from $x_1$ on the cycle is equal to the probability of finding the walker at position $x_2\ominus x_1$ starting from the position  $x_0=0$.
It follows that $\mu(x)$ represents the transition probability  between
two sites  separated by a displacement $x$ (modulo $N$).

\begin{figure}[t]
    \centering
\includegraphics[width=0.5\linewidth]{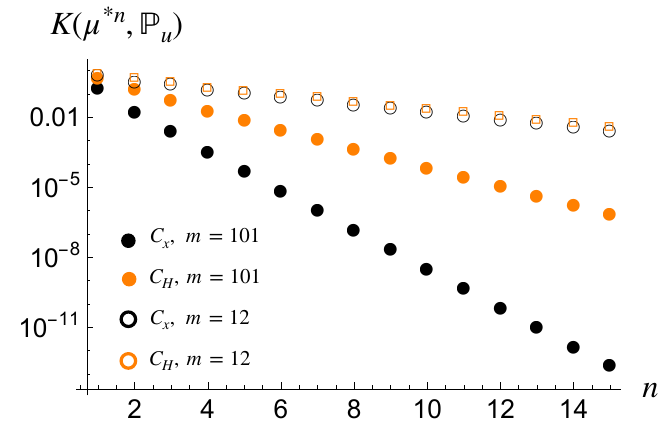}
    \caption{{Cycle graph size: $N=25$. Kolmogorov distance between the convoluted probability $\mu^{*n}(x)$ and the uniform distribution as a function of the number of iterations  $n$, for two different  coin operators, $\coinoperator_H$ and $\coinoperator_X$ and two different numbers of timestep $m$. The initial coin state  is  fixed: $\ket{c_0}=\frac{1}{\sqrt{2}}(\ket{\uparrow}+\ket{\downarrow})$.}}
    \label{fig:convolution}
\end{figure}

The marginal probability of $X_1$ is thus given by $ \probdistr(X_1) = \mu(x_1).$
For the second step,  the chain rule for  joint probabilities gives:
\begin{equation}
\begin{split}
    \probdistr(X_{1:2}) 
    = \mu(x_2 \ominus x_1) \mu(x_1),
\end{split}
\end{equation}
and, marginalising over $X_1$, we obtain
\begin{equation}
    \probdistr(X_2) = \sum_{x_j=0}^{N-1} \mu(x_2\ominus x_j) \mu(x_j) = \mu^{\star 2}(x_2),
\end{equation}
where ``$\star$'' denotes the convolution operator, defined recursively as
\begin{equation}
    \mu^{\star n}(x) = \left[\mu\star\mu^{\star(n-1)}\right](x) \qquad \text{ for } n \geq 2,
\end{equation}
with $\mu^{\star 1}(x) = \mu(x)$. 
Generalizing, we obtain the marginal probability for the $n^{th}$ extraction:
\begin{equation}
\probdistr(X_n)=\mu^{\star n}(x_n).
\end{equation}
Proving that the marginal probability $\probdistr(X_n)$ converges to the uniform is, hence, equivalent to proving that
\begin{equation}
    \lim_{n\to\infty} \mu^{\star n}(x) = \frac{1}{N} \qquad  \forall x\in [0,N-1].
\end{equation}
A {classical} random walk that satisfies this property is called \textit{ergodic}. 
We now resort to the following theorem for random walks on finite groups \cite{Diaconis_1988,McCarthy21}:
\begin{theorem}[Ergodic theorem]
    A 
    random walk with transition probability $\mu$ on the $N$-cycle is ergodic if and only if the support of $\mu$ is not contained in any coset of a proper subgroup of the $N$-cycle.
\end{theorem}
As long as the number {of nodes of graph} are properly chosen, the hypotheses of the ergodic theorem are easily satisfied. 
In fact, every {proper subgroup} of the $N$-cycle is generated by each of the non-trivial divisors of $N$. As an example, consider the group $C_6 = \{0,1,2,3,4,5\}$, which is isomorphic to the additive group of $\mathbb{Z}_6$. It has two proper subgroups, namely $\{0,2,4\}$, generated by 2, and $\{0,3\}$, generated by 3. 
The cosets of a subgroup are all the possible translations of such a subgroup on the cycle. 
To ensure an ergodic random walk, it is therefore sufficient that the support of $\mu(x)$  does not follow the structure of any subgroup (or its cosets). This implies, for instance, that when { the number of nodes 
$N$ is even}, the conditions of the ergodic theorem are not satisfied {since the support of $\mu$ is always contained in a coset of a proper subgroup of the $N$-cycle}, thus preventing  convergence to the uniform distribution.
 \begin{figure}[t]
    \centering
\includegraphics[width=0.5\linewidth]{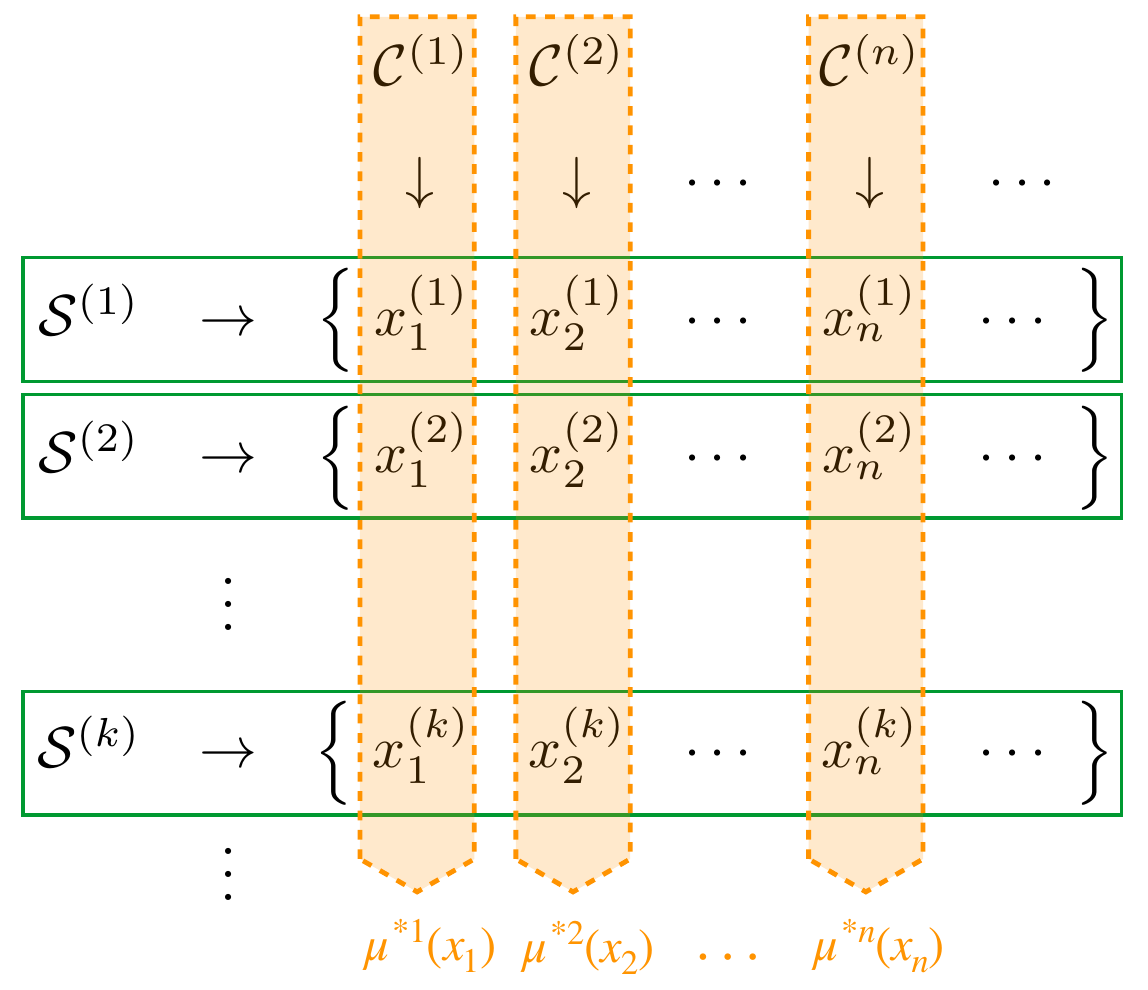}
 \caption{
{A single application of the protocol of Sec.~\ref{sect:sampling} yields a  string $\mathcal{S}^{(k)}$ of correlated outcomes, although each element is marginally distributed according to $\mu^{*n}$, which converges to the uniform distribution for $n\gg1$. By resetting and  repeating  the protocol  to generate independent strings $\{\mathcal{S}^{(k)}\}$,  the set $\mathcal{C}^{(n)}=\{x_n^{(1)},x_n^{(2)},\dots\}$ forms a sample of iid uniformly distributed variables. }}
    \label{fig:scheme}
\end{figure}
\newline
\newline
{
Thanks to the ergodic theorem, the $n^{\rm{th}}$ extracted value $x_n$ is distributed according to $\mu^{*n}$ which converges to the uniform distribution for sufficiently large  $n$. However, successive outcomes within a single realization of the protocol i.e., steps I-IV, are correlated and therefore cannot be regarded as independent random samples. 
To generate an ensemble of independent and uniformly distributed outcomes, the protocol described in Section~\ref{sect:sampling} must be repeated  multiple times. Specifically, one  generates
 independent strings 
 $\{\mathcal{S}^{(k)}=X_{1:n}^{(k)}\}$, by restarting the protocol from step I after  each string realization. Then, from each string only the final outcome $x_n^{(k)}$ is retained. The resulting set $\mathcal{C}^{(n)}=\{x_n^{(1)},x_n^{(2)},\dots\}$  then constitutes an independent sample drawn from the uniform distribution. A schematic representation of this procedure is shown in Fig.~\ref{fig:scheme}.}
 \newline
{Different choices of $m$, $\coinoperator$ and initial state, for fixed $N$, affect the  convergence speed of  the marginal distribution toward uniformity. The convergence can be quantified through the Kolmogorov distance between the marginal $\mu^{*n}$ and the uniform  distribution $\mathbb{P}_u$, namely:
 \begin{equation}\label{kolmogorov:dist}
 K(\mu^{*n},\probdistr_u)=\frac12 \sum_{x=0}^{N-1}\,\left|\mu^{*n}(x)-\frac{1}{N}\right|.
 \end{equation}
 As an illustrative example, in Fig.~\ref{fig:convolution},  we report  $K(\mu^{*n},\probdistr_u)$ as a function of $n$, for two different coin operators and number of the DTQW steps $m$, for a fixed initial state and graph size $N$.  Faster convergence toward uniformity is observed when the  transition probability  is more evenly distributed over the nodes of the graph. The  probability distributions corresponding to $m=12$ and $m=101$ for the coin operators used in Fig. \ref{fig:convolution} are shown in the the bottom panels of Fig.~\ref{fig:transition_probabilities}.
It is worth noting that, for a cycle graph, which has a finite a periodic structure, the spatial distribution $\mu$  spreads monotonically only for small values of $m$ while, as $m$ increases further, the spreading exhibits a non-monotonic behavior.
These considerations shows that the choice of different coin operators and the  parameter $m$ can significantly influence the convergence rate, indicating that appropriate operating  conditions are required to achieve almost-uniform sampling.}
\newline
We remark  that the same analysis can also be applied  to other choices of the coin operator and initial coin states, provided that the  conditions of the ergodic theorem are satisfied. 
\newline\newline
Assuming that the hypotheses of the ergodic theorem are satisfied, it is also possible to derive a lower bound for the convergence of the Shannon entropy of the marginal distribution to its maximum value, which corresponds to   the uniform distribution. Given the Fourier coefficients of $\mu$ as
\begin{equation}
    \hat{\mu}(k) = \sum_{s=0}^{N-1} \mu(s) e^{-i 2\pi ks/N},
\end{equation}
the Diaconis--Shashahani bound~\cite{Diaconis_1988} yields for the Shannon entropy of the marginal distributions:
\begin{equation}
    H(\mu^{\star n}) \geq \log_2 N - \left(\log_2 N + 1\right)\sqrt{\sum_{k=1}^{N-1}|\hat{\mu}(k)|^{2n}}.
\end{equation}
If the support of $\mu$ is not contained in any coset of a proper subgroup of $\mathbb{Z}_N$, 
then every nonzero Fourier coefficient satisfies $|\hat{\mu}(k)|<1$. 
Under this hypothesis, the Diaconis–Shahshahani bound ensures that 
$H(\mu^{\star n}) \to \log_2 N$ as  $n\to\infty$,  since the sum under the square root tends to zero exponentially fast. 
{\subsection{Single application of the protocol}
It is important to remark that a single run of the protocol does not provide a sample of independent random variables. Indeed, the values within a single string are generated  by a classical Markov chain with  the same transition probability $\mu$ and are therefore correlated. As a consequence, despite each outcome being marginally uniformly distributed in the asymptotic regime, a single string cannot be regarded as a sample of true randomness in the sense of independent and identically distributed variables.}\\
In general, uniform marginal distributions do not necessarily imply  a uniform joint distribution.
The dependence between the two random variables $X_n$ and $X_{n-1}$  is strongest when the conditional distribution $\probdistr(X_n|X_{n-1})$ deviates most from the uniform distribution $\probdistr_u(X_n)$. 
Therefore, to minimize correlations between consecutive outcomes, the {protocol} parameters 
{$(m,\coinoperator,\ket{c_0})$, for a fixed graph size,} should be chosen so that the transition probability,  $\mu(x_n\ominus x_{n-1})$ is as close as possible to uniform. {This can be quantified by minimizing the Kolmogorov distance between the transition probability and the uniform distribution, $K(\mu,\probdistr_u)$. }

{As an  example, Fig.~\ref{fig:transition_probabilities} shows the Kolmogorov distance as a function of the number of timesteps $m$, for a fixed graph size $N$ and initial coin state, for two representative  coin operators. We consider $C_X$ which leads to a symmetric transition probability, and  $C_H$ which results in an asymmetric dynamics for the chosen initial  state (see bottom plots). Small values of the Kolmogorov distance indicate a  transition probability closer to uniformity and therefore weaker correlations between following outputs. 
\begin{figure}[t]
    \centering
\includegraphics[width=0.6\linewidth]{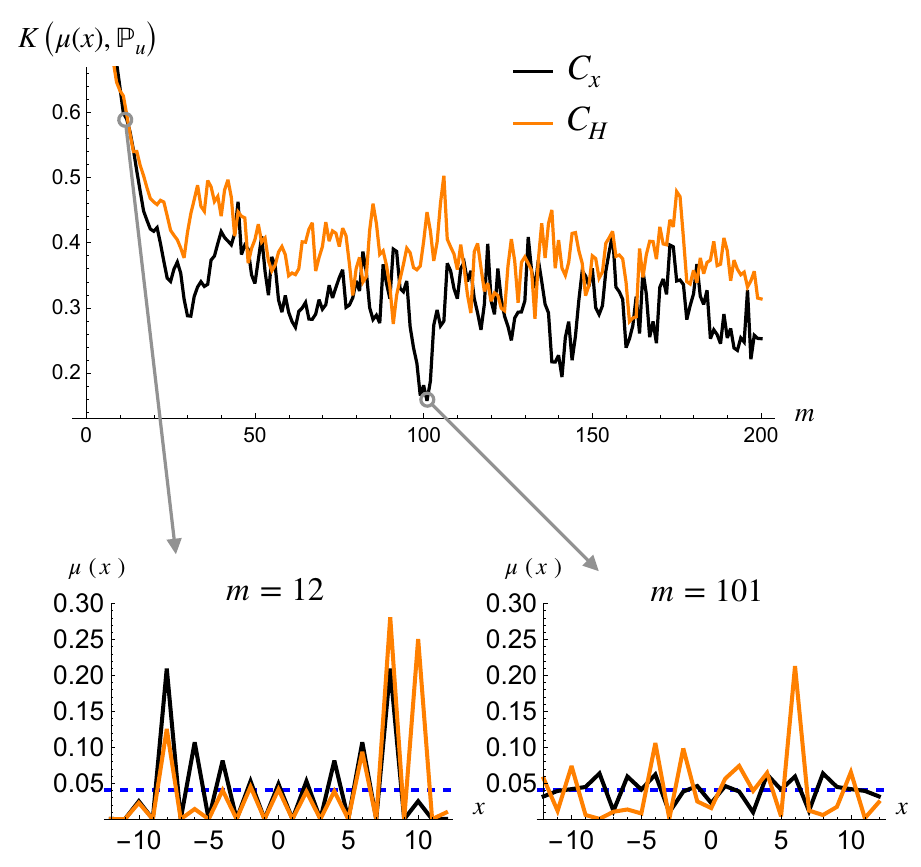}
    \caption{Cycle graph size: $N=25$. Upper plot: Kolmogorov distance between the transition probability $\mu(x)$ and the uniform distributions as a function of the number of timesteps $m$, for two different  coin operators, $\coinoperator_H$ and $\coinoperator_X$. The initial coin state for both evolutions is $\ket{c_0}=\frac{1}{\sqrt{2}}(\ket{\uparrow}+\ket{\downarrow})$. Bottom plots:  
    Transition probability $\mu(x)$  for two selected values of  $m$, obtained for the same coin operators and initial state as above. The blue dashed line is a guide for the eye, corresponding to the uniform distribution.}
    \label{fig:transition_probabilities}
\end{figure}
For  small  $m$, $K$ decreases as the transition probability spreads over the nodes. For larger $m$, however, the dependence on $m$ becomes non-monotonic, showing that increasing the number of steps does not necessary improve uniformity. Consequently, the choice of $m$ must be made carefully to mitigate correlations between successive outcomes. Similar considerations apply to other choices of coin operators and initial conditions.}
\newline

{The correlation length in a single string can be quantified through the correlation function:
\begin{align}
    C(r)=\langle x_j x_{j+r} \rangle-\langle x_j \rangle\langle  x_{j+r} \rangle.
\end{align}
Correlations decay for outcome separations $r\gtrsim \overline{n}$, where $\overline{n}$ is the minimum number of iterations required for the the random walk  to become ergodic, i.e. such that $\mu^{*\overline{n}}(x)\approx 1/N$, (see Appendix~\ref{corrLength} for details).
}

\section{Conclusions}\label{sect:Conclusions}
Being able to sample uniformly distributed numbers  is a crucial task for many applications, from quantum cryptography to quantum information processing. 
In this work, we have proposed and analyzed a quantum-inspired protocol for sampling  from an almost–uniform distribution by exploiting the dynamics of a discrete–time quantum walk on the cycle graph, under the assumption that no external source of randomness is available. 
To achieve randomness in this setting, it is necessary to introduce irreversibility into the procedure. We do so by means of a measure–and–reset scheme:
 the quantum walk is initialized in a state that gives rise to the transition probability $\mu(x)$ with a  significant support across the nodes of the graph;
the walker evolves for $m$ steps, after which its position is measured; the corresponding vertex label is recorded;   the walker is  reset to the measured site with its coin prepared in the fixed initial state.  The previous steps are repeated $n$ times. In the limit of large $n$, this protocol returns a string of $n$ integers with almost-uniform randomness.

Although correlations between successive outcomes cannot be completely removed, they can be significantly reduced by a suitable choice of the timestep parameter $m$, for fixed initial state and coin operator. 
Using  the ergodic theorem for Markov chains on the $N$-cycle and  the properties of iterated self-convolution of  transition distributions, we have proved that, for sufficiently long strings, the marginal distribution $\probdistr(X_n)$ converges to the uniform distribution in the limit of large $n$. 
Moreover, by analyzing the Fourier spectrum of the transition probability $\mu(x)$, we have shown that the convergence of the Shannon entropy towards its maximal (uniform) value can be quantitatively bounded using the Diaconis–Shahshahani inequality.

The results of our research can stimulate further investigations on randomness generation based on quantum walk evolutions.  This simple model may serve as a building block for more complex architectures and for  the design of realistic and reliable platforms for the quantum random number generation. Such platforms could to be embedded into larger and more sophisticated quantum devices, such as quantum computers or quantum networks.

\ack{
The authors thank Andrea Smirne for useful discussions. CB and SO acknowledges support from MUR and EU
via the PRIN 2022 scorrimento Project EQWALITY (Contract
N. 202224BTFZ), QSTI-Spoke1-BaC QSynKrono (contract n.~PE00000002-QuSynKrono), and Piano di Sviluppo UniMi 2023.
MR acknowledges funding by the Irish Research Council under Government of Ireland Postgraduate Scheme grant N. GOIPG/2022/2321. This work was supported by RIT (Research IT, Trinity College Dublin), some of the calculations were performed on the Lonsdale cluster maintained by the Trinity Centre for High Performance Computing. This cluster was funded through grants from Science Foundation Ireland.}

\appendix

\section{On the Cesàro convergence for a DTQW distribution with a degenerate spectrum}
\label{app:Cesàro_convergence}
Here we derive  the long-time averaged (Cesàro) spatial probability distribution for a discrete-time quantum walk with  a degenerate spectrum and show that, in general, it does not converge to the uniform one. 

Following Ref.~\cite{Aharonov_2001}, we consider a walk with initial state $\ket{\Psi_0}=\ket{c_0}\otimes\ket{0}$. The state after $t$ steps is given by $\ket{\Psi_t}=\hat{\mathcal{U}}^t \ket{\Psi_0}$, where $\hat{\mathcal{U}}$ is the unitary step operator of Eq.~\eqref{stepoperator}. 
The time-averaged probability  distribution of a walker over the node set $\{\ket{v}\}$ is
\begin{equation}
\overline{p}_T(v) = \frac{1}{T} \sum_{t=0}^{T-1} p_t (v)
\end{equation}
The long-time limit averaged distribution is
 \begin{align}
     \pi(v) = \lim_{T\to\infty} \overline{p}_T(v).
 \end{align} 
To compute $\pi(v)$, we consider the following theorem.
\begin{theorem}
    Let $\{\ \lambda_j,\ket{\phi_j}\}$ be the eigenvalues and eigenvectors of  the  step operator $\hat{\mathcal{U}}$.  If $a_n=\braket{\phi_n}{\Psi_0}$, then the limiting time-averaged distribution is
    \begin{equation}
        \pi(v) = \sum_{c}\sum_{n,m} a_n a_m^* \braket{c,v}{\phi_n}\braket{\phi_m}{c,v}\delta_{\lambda_n\lambda_m},
        \label{a2}
    \end{equation}
where  $\ket{c,v}$ denotes the DTQW basis state in $\coinspace\otimes\positionspace$. Notice that the sum has contributions only from the pairs $(n,m)$ such that $\lambda_n = \lambda_m$.
\begin{proof}
Upon expanding the generic state $\ket{\Psi_t}$ of the DTQW after $t$ time steps in the basis state $\ket{c,v}$, we have:
        \begin{align}
           \vert \braket{c,v}{\Psi_t} \vert^2 &=  \left\vert \sum_n \lambda_n^t a_n \braket{c,v}{\phi_n}\right\vert^2 = \sum_{n,m} a_n a_m^* \,(\lambda_n\lambda_m^*)^t \braket{c,v}{\phi_n}\braket{\phi_m}{c,v}.
            \label{a3}
        \end{align}
Now, as proved in Ref.~\cite{Aharonov_2001}, for $T\gg 1$ the quantity $ T^{-1}\sum_{t=0}^{T-1} (\lambda_n\lambda_m^*)^t$ 
converges to 1 if $\lambda_n=\lambda_m$ and to 0 otherwise. Thereafter,
taking the time average
\begin{equation}
\frac{1}{T}\sum_{t=0}^{T-1} \vert \braket{c,v}{\Psi_t} \vert^2
\end{equation}
and summing over all the coin states $\ket{c}$, one straightforwardly obtains Eq.~\eqref{a2}. 
    \end{proof}
\end{theorem}

We can now prove the following Corollary.

\begin{corollary}
    Let $\hat{\mathcal{U}}$ be the step operator of a DTQW on an $N$-cycle,   with  coin operator $\coinoperator_X$ defined in Sec. \ref{sect:Discrete_time_quantum_walks}. If $N$ is odd, then
 the limiting time-averaged distribution Eq.~(\ref{a2}) does not converge to the uniform one.

\begin{proof}
For a walk on a cycle, $\hat{\mathcal{U}}$ has the form given in Eq.~\eqref{stepoperator}, with operators $\coinoperator$ and $\shiftoperator$ defined in the main text.  In order to  build the eigenvectors of such operator, we introduce  the Fourier (momentum) states 
\begin{equation}
\ket{\chi_k} = \frac{1}{\sqrt{N}} \sum_{v=0}^{N-1} e^{i\frac{2\pi}{N}kv} \ket{v},
            \label{eq:characters_state}
\end{equation}
with $k=0,\dots,N-1$.
A reasonable guess for the form of the $2N$ eigenvectors of $\hat{\mathcal{U}}$ is then:
\begin{align}
    \label{eq:form_eigenstates_cycle} \hat{\mathcal{U}}\ket{\phi_k^{\pm}}&=\lambda_k^{\pm}\ket{\phi_k^{\pm}}
\end{align}
with
\begin{align}
\ket{\phi_k^{\pm}}&=   \ket{\gamma_k^{\pm}} \otimes \ket{\chi_k},\label{autostati}
\end{align}
where  $\lambda_k^{\pm} = e^{\pm i \theta_k}$  and  $\ket{\gamma_k^{\pm}}$  are  eigenvectors of the $2\times 2$ matrix
\begin{equation*}
H_k = \Lambda_k \coinoperator,\qquad \Lambda_k=\begin{pmatrix}
    e^{2\pi ik/N}&0\\
    0&e^{-2\pi ik/N}
\end{pmatrix}
\end{equation*}
with eigenvalues $\lambda_k^{\pm} $. We remark that $\braket{\gamma_k^s}{\gamma_k^{s'}}=\delta_{ss'}$ and, in general, 
for $k\neq j$, {$\braket{\gamma_k^s}{\gamma_j^{s}}\neq0$}, with $s,s'\in \{\pm\}$.
Using the expression of the coin operator of Eq.~\eqref{eq:coin_operator}, we obtain the eigenvalue equation for  $H_k$:
    \begin{equation}
        (\lambda^{\pm }_k)^2 - \sqrt{2}\,\lambda_k^{\pm}\,\cos\left( \frac{2\pi k }{N}\right) + 1 = 0.
    \end{equation}
 which gives the relation:
 \begin{equation}
        \cos(\theta_k) = \frac{1}{\sqrt{2}}\cos\left( \frac{2\pi k}{N}\right).
\end{equation}
For a fixed value of $k$, there are two distinct solutions associated with 
 $\theta_k$ e  $2\pi-\theta_k$ belonging to the intervals $[\pi/4, 3\pi/4]$ and $[ 5\pi/4,7\pi/4]$.
However, for different momenta $k$ and $k'$ the  degeneracy occurs when  
  $k'=N-k$. 
Referring to Eq.~\eqref{a2}, the only terms that survive in the sum are those corresponding to degenerate eigenvalues and those with $n=m$. 

If we consider $N$ odd, all eigenvectors are pairwise degenerate except for  $k=0$ which corresponds to the only non-degenerate eigenvectors $\ket{\phi_{k=0}^{\pm}}$.
Substituting Eq.~\eqref{autostati} into the long-time averaged distribution $\pi(v)$ and  summing over the coin states,
 one obtains:
 \begin{align}
\pi(v)&= \frac{1}{N} +
 \frac{1}{N^2}\sum_{n=1}^{N-1}\exp\left(-i\frac{4\pi n v}{N}\right)\times\sum_{s=\pm}\braket{\gamma_n^s}{\gamma_{N-n}^s}
  \bra {\gamma^s_{N-n}}\Pi_0\ket{\gamma_n^s} 
  \label{pi}
\end{align}
where $\Pi_0=\ketbra{c_0}{c_0}$ is the projector onto the initial coin state. 
The first term, $N^{-1}$, is the contribution from same index eigenvectors, i.e., $n=m$ in Eq.~(\ref{pi}) and yields a uniform distribution: this is  the only term that survives in the presence of non-degenerate spectra. The second term in Eq.~(\ref{pi}) does not vanishes and induces deviation from uniformity, hence the thesis follows.
\end{proof}
\end{corollary}

Similar arguments can be applied to the case with an even number of nodes $N$. We eventually note that, in the presence of the Hadamard coin, the last term in Eq.~(\ref{pi}) vanishes leading to the convergence of the Cesàro distribution to the uniform one \cite{Aharonov_2001}.

{
\section{Correlation length}
\label{corrLength}
Given the string:
\begin{equation}
{\cal S} = \{x_1,\ldots,x_j,\ldots,x_{j+r},\ldots\}
\end{equation}
of numbers $x_{k} \in \{0,1,2,\ldots,N-1\}$ generated by the protocol in Sec. \ref{sect:sampling}, we introduce the correlation function:
\begin{equation}
C(r) = \langle x_j \, x_{j+r}\rangle-
\langle x_j \rangle \langle x_{j+r} \rangle,
\end{equation}
where, following the formalism of Subsection~\ref{ss:convergence}:
\begin{equation}
\langle x_j \rangle = \sum_{x_j} x_j\,\mustar{j}{x_j}
\end{equation}
with:
\begin{align}
\mustar{j}{x_j} &= \sum_{x_1}\cdots\sum_{x_{j-1}} \prod_{k=1}^{j-1}\mu(x_{k+1}\ominus x_k)\, \mu(x_1\ominus x_0),\\
&\equiv \mustar{j}{x_j|x_0} \label{app:step:one}
\end{align}
$x_0$ being a {\it fixed} starting number (see Subsection~\ref{ss:convergence}). Thereafter, we have:
\begin{align}\label{corr:j:jp1}
&\langle x_j \,x_{j+r}\rangle =\sum_{x_j}\sum_{x_{j+r}} x_j\, x_{j+r}\,
\mustar{(j+r)}{x_{j+r}|x_j}\, \mustar{j}{x_j|x_0}\,,
\end{align}
with:
\begin{align}
&\mustar{(j+r)}{x_{j+r}|x_j} =\sum_{x_{j+1}}\cdots\sum_{x_{j+r-1}}
\prod_{k=j+1}^{j+r-1}\mu(x_{k+1}\ominus x_k)\, \mu(x_{j+1}\ominus x_j)
\end{align}
We can now estimate the correlation length, that is the minimum value of $r$ such that $C(r)$ vanishes. Assessing the Kolmogorov distance (\ref{kolmogorov:dist}) between $\mustar{j}{x_j|x_0}$ and the uniform distribution $\mathbb{P}_u$, and invoking the ergodic theorem (see Subsection~\ref{ss:convergence}), we can find the number of iterations $\overline n$ such that $\mustar{j}{x_j|x_0} \approx 1/N$. If $j \gtrsim  {\overline n}$, the Eq.~(\ref{corr:j:jp1}) reduces to:
\begin{align}\label{corr:j:jp}
\langle x_j x_{j+r}\rangle \approx \frac1N \sum_{x_j}\sum_{x_{j+r}} x_j x_{j+r}
\mustar{(j+r)}{x_{j+r}|x_j}\,,
\end{align}
while:
\begin{equation}
\langle x_j \rangle = \langle x_{j+r} \rangle
\approx \frac{N-1}{2}.
\end{equation}
We note that, due to the ergodic theorem, the $x_j$'s appearing in $\mustar{(j+r)}{x_{j+r}|x_j}$ of Eq.~(\ref{corr:j:jp}) are uniformly distributed and, thus, given $j \gtrsim  {\overline n}$ and for a fixed $x_j$, $\mustar{(j+r)}{x_{j+r}|x_j}$ plays the role of (\ref{app:step:one}) with the replacement $x_0 \to x_j$ (the new starting number) and $x_j \to x_{j+r}$ (the drawn number after $r$ steps). It is straightforward to see that if also $r \gtrsim {\overline n}$, then $\mustar{(j+r)}{x_{j+r}|x_j} \approx 1/N$, and Eq.~(\ref{corr:j:jp}) reduces to:
\begin{align}
\langle x_j x_{j+r}\rangle \approx \left( \frac1N\sum_{x_j}x_j\right) \!\!\left(\frac1N\sum_{x_{j+r}} x_{j+r}
\right) = \langle x_j \rangle \langle x_{j+r} \rangle\,,
\end{align}
leading to $C(r) \approx 0$. Clearly, the actual value of $\overline n$ depends on the transition probability and, thus, on the features of the quantum walk on the cycle. We can estimate it posing a suitable threshold on the Kolmogorov distance which is monotone with respect to the the number of iterations of our algorithm (see Fig.~\ref{fig:convolution}).
}

\bibliography{bibliography}
\end{document}